# Doping dependent vortex activation energy and pseudogap in Y123

**S H Naqib and R S Islam**

*Department of Physics, University of Rajshahi, Rajshahi-6205, Bangladesh*

E-mail: salehnaqib@yahoo.com

## Abstract

The temperature and magnetic field dependent activation energy, $U(T, H)$, is one of the most important parameter in the field of applied superconductivity as it primarily determines both the crtical current density and the irreversibility field. Previously, we have determined the doping dependent $U(T, H)$ from the analysis of field dependent resistive transitions in high-quality *c*-axis oriented crytalline thin films of Y123 (arXiv:1207.4312). In this short communication, we have showed a direct link between the characteristic field $H_0$ that sets the magnitude of $U(T, H)$ and the pseudogap temperature, $T^*$. The strong dependence of $H_0$ on the in-plane hole content, $p$, seems to follow from the *p*-dependent evolution of the pseudogap energy scale ($T^*$, when expressed in temperature) which reduces the superconducting condensation energy as hole concentration decreases.

## 1. Introduction

What determines the flux activation energy in high-$T_c$ cuprates? This is a question of significant importance since activation energy determines the magnitude as well as temperature and field dependences of the critical current density and irreversible magnetic field [1 − 3]. It is believed that the vortex dynamics of cuprate superconductors are complicated because of competing roles of dimensionality, large thermal fluctuations associated with high-$T_c$, and quenched disorder. In the mean-field region (where non-Gaussian critical fluctuations are absent) the resistive broadening of the field dependent resistivity data, $\rho(T, H)$, is usually studied within the thermally assisted flux flow (TAFF) formalism [1 – 9].

In some earlier investigations we have shown that the *ab*-plane critical current density is dominated by the depairing contribution, shows the same qualitative trend as shown by the *p*-dependent superfluid density [10 – 12] and therefore, the phase stiffness of the superconducting (SC) order parameter. In this short communication, we have shown that there is a clear correspondence between the doping dependent values of $H_0$ and the characteristic pseudogap (PG) temperature $T^*$.

## 2. Experimental samples

Crystalline *c*-axis oriented thin films of $YBa_2Cu_3O_{7-\delta}$ were used for the in-plane resistivity, $\rho_{ab}(T, H)$, measurements. Thickness of the films lies within the range (2800 ± 300) Å. Details of sample preparation, characterization, and resistivity measurements can be found in refs. [9 – 14]. The hole concentrations were changed by changing the oxygen deficiency in the $CuO_{1-\delta}$ chains by annealing the films under different temperatures in different oxygen partial pressures. The in-plane room-temperature thermopower, $S_{ab}[290\ K]$ was used to calculate the planar hole concentration following the procedure proposed by Obertelli *et al.* [15].

## 3. Experimental results and analysis

The *ab*-plane resistivity data under an applied magnetic field along the crystallographic *c*-direction is shown in Fig. 1 [9]. The resistive transition was investigated using the TAFF scenario [9]. Where the vortex activation energy was found to follow the functional form given by $U(T, H) = (1-t)^m (H_0/H)^{-\beta}$, where $t = T/T_c$, is the reduced temperature and $H_0$, is a field scale that determines the magnitude of the activation energy for a given sample composition at fixed magnetic field and temperature [8, 9]. The analysis presented in ref. [9], revealed the evolution of the temperature exponent, $m$, field exponent, $\beta$, and $H_0$ as *p*-values are varied. We reproduce the $H_0(p)$ behavior in Fig. 2 [9].

In this study, the characteristic PG temperature scale has been located at the onset of the downward deviation of experimental the $\rho(T)$ data from its high-$T$ linear fit. Zero-field resistivity data is used here. It should be mentioned that application of magnetic field suppresses $T_c$ but has a minimal effect on $T^*$ [16]. This method is illustrated in Fig. 3. As the PG reveals itself, the available low-energy quasiparticle (QP) states near the chemical potential get progressively depleted and this reduces the QP scattering rate, thereby reducing the resistivity at high-temperatures much above the superconducting transition. The extracted values of $T^*(p)$ are in excellent agreement with those found in previous studies [17 – 21]. Fig. 4 exhibits the plot of $T^*(p)$ versus $H_0(p)$. An almost linear anti-correlation is found between these two apparently unrelated parameters.

## 4. Discussion and conclusions

As can be seen from Fig. 2, $H_0$ decreases rapidly in the underdoped (UD) region and seems to peak in the overdoped (OD) side. It is interesting to note that even though the $T_c$ values are almost identical for the optimally (OPD) and the OD samples, $H_0$ is substantially higher for the OD compound. The OPD compound is relatively more disordered due to oxygen deficient $CuO_{1-\delta}$ chains, whereas the OD one has a higher superfluid density due to a smaller PG in the low-energy electronic density of states [11, 22]. These indicate that the role played by superfluid density/SC condensation energy is significantly greater in enhancing the pinning potential than that due to the oxygen defects in the $CuO_{1-\delta}$ chains even if they can act as additional pinning sites.

It is reasonable to assume that the flux line is pinned at a site where the SC order parameter is partially or almost completely suppressed. In this situation, the energy of the vortex core would manifest itself as the energy barrier for motion of the flux line and therefore, would be equal to the activation energy $U_0$ [23]. Here $U_0$ is the activation energy at zero temperature. The energy density at the vortex core varies as $H_c^2$, at the same time the SC condensation energy, $U_{sc}$, also varies as $H_c^2$ ($H_c$ is the thermodynamic critical field). Thus, $U_0 \sim U_{sc} \sim H_c^2$. At this point it is possible to establish a direct relation between the electronic energy density of states (EDOS) and activation energy by recalling the expression for SC condensation energy given by $U_{sc} \sim \Delta_{sc}^2 N(\varepsilon_F) \sim U_0$, where $\Delta_{sc}$ is the coherence gap amplitude and $N(\varepsilon_F)$ is the EDOS at the Fermi level. Since $\Delta_{sc}$ does not vary with decreasing hole content for samples with a finite PG [24], and PG progressively depletes $N(\varepsilon_F)$ with underdoping, the above expression directly links $U_0(p)$ with $T^*(p)$. This scaling analysis also offers an explanation for the observed anti-correlation between $T^*(p)$ and $H_0(p)$, at least at a qualitative level, since this characteristic field sets the magnitude for $U_0(p)$ at any finite applied magnetic field below the superconducting transition temperature.

Therefore, in order to maximize the critical current density or the irreversibility field (both of which depend directly on activation energy for flux lines/bundles), one must control the sample composition so that the PG is small or zero, without reducing $T_c$ too much by gross overdoping. In the previous work [9] we noted that the 1/8[th] anomaly (where the spin/charge stripe correlations are at their strongest) [25 – 27] has no visible effect on $H_0(p)$. This supports the proposal made in refs. [26, 27], that PG and spin/charge density orders are not directly related phenomena. In addition, it indicates that the inhomogeneity in the charge/spin arrangements induced by the incipient stripe instability does not have a significant effect on the vortex dynamics in Y123.

## Acknowledgements

The authors would also like to thank the Abdus Salam International Centre for Theoretical Physics (AS-ICTP), Trieste, Italy, for the hospitality.

**Figure captions**

Figure 1 (color online): The field dependent *ab*-plane resistivity of (a) $p = 0.170$, (b) $p = 0.150$, (c) $p = 0.135$, and (d) $p = 0.118$ thin films under different magnetic fields applied along the crystallographic *c*-direction [9]. The hole contents are accurate within $\pm\ 0.005$.

Figure 2 (color online): The characteristic field scale $H_0$ versus hole concentration. The dashed line is a second-order polynomial fit to the $H_0(p)$ data [9].

Figure 3 (color online): Extraction of the characteristic PG temperature, $T^*$, from the resistivity data. The black dashed lines are the high-$T$ linear fits to the $\rho_{ab}(T)$ data. Arrows mark the onset of the downward deviations of the experimental data from the linear fits, therefore, $T^*$.

Figure 4 (color online): $T^*(p)$ versus $H_0(p)$. An anti-correlation is seen. The thick dashed straight line is drawn as a guide for the eye.

Figure 1

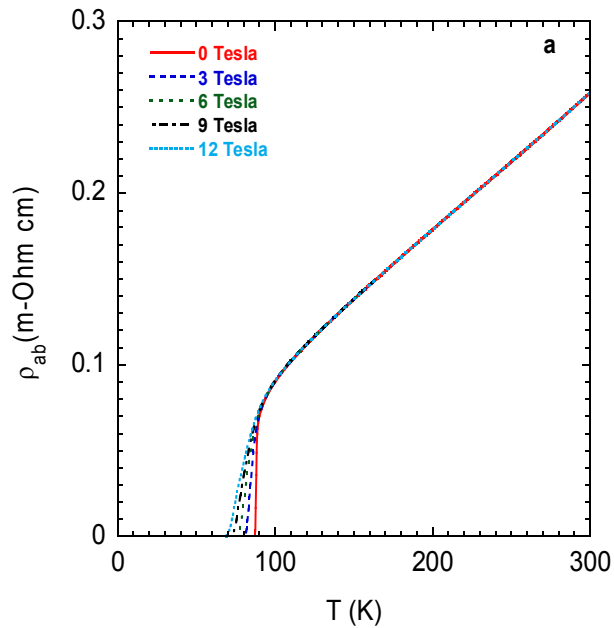

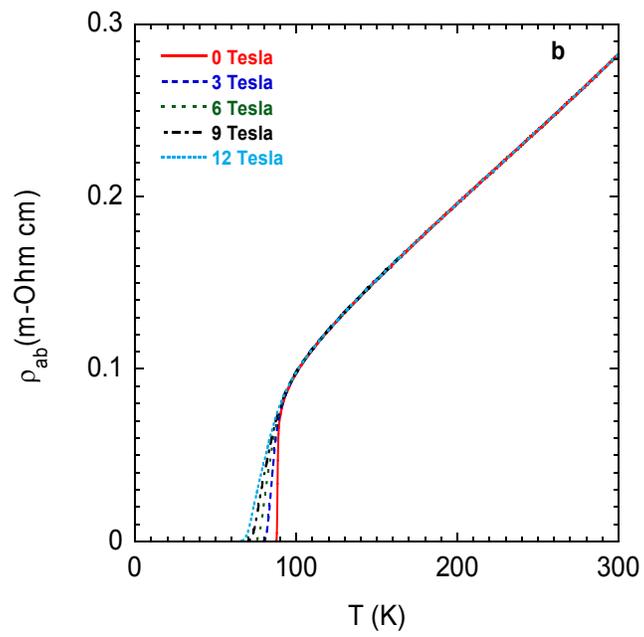

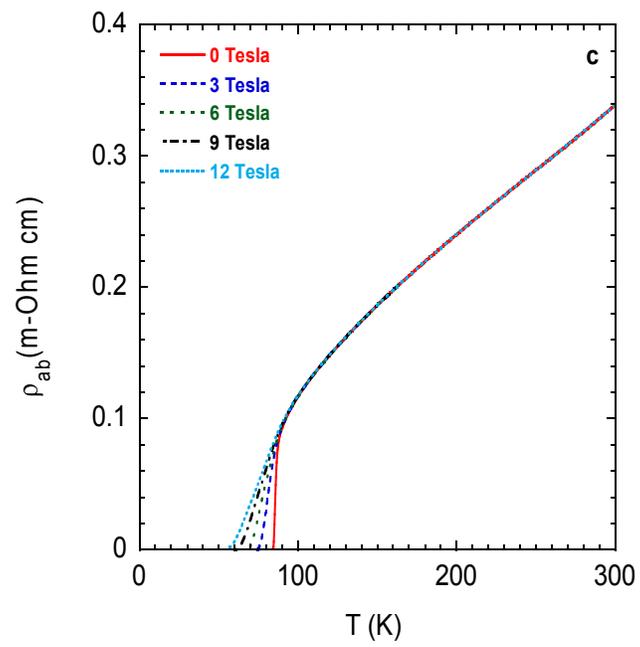

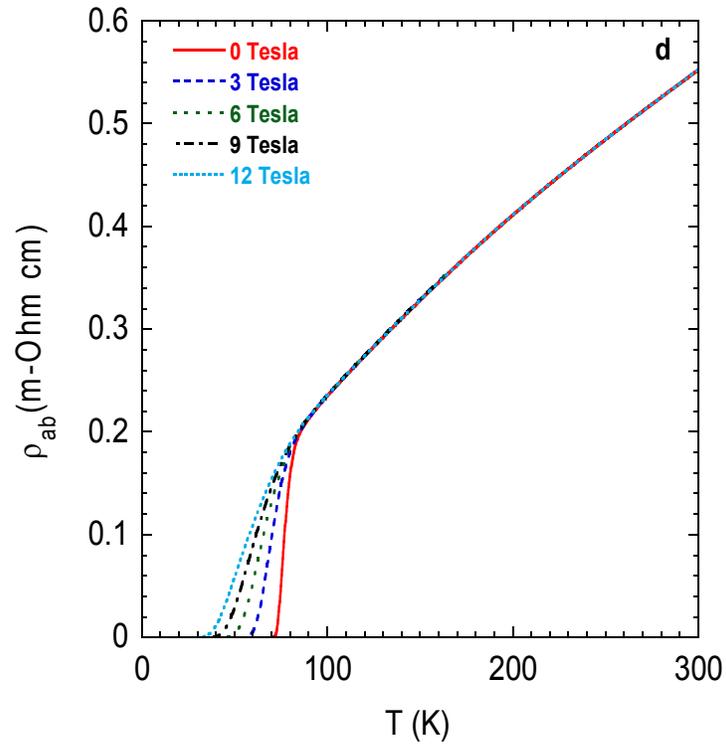

Figure 2

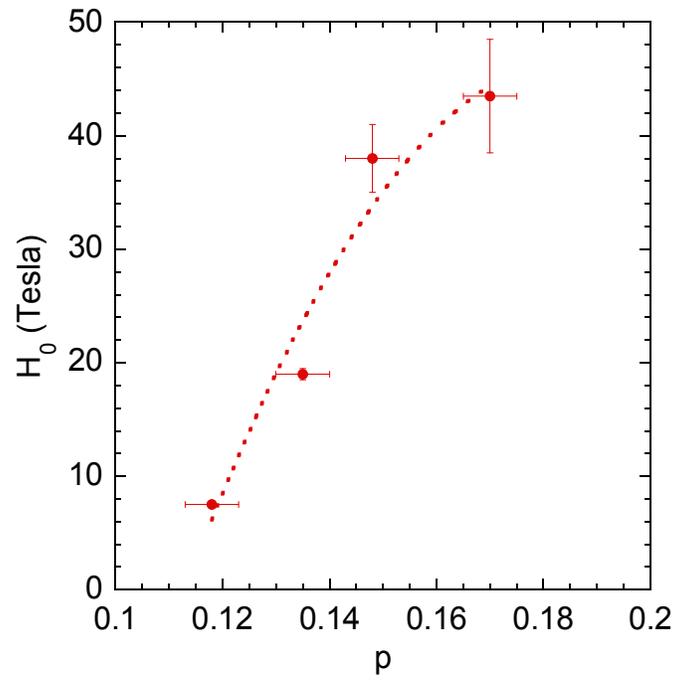

Figure 3

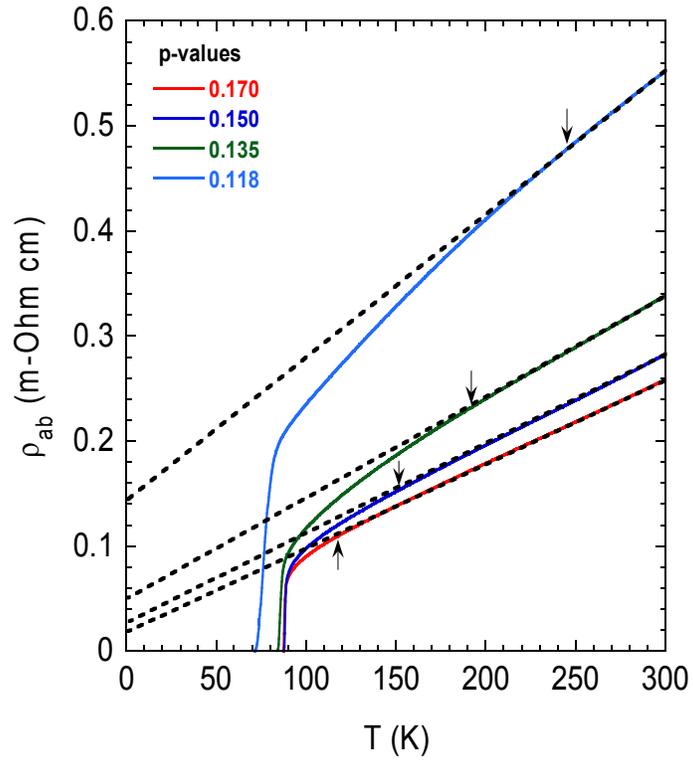

Figure 4

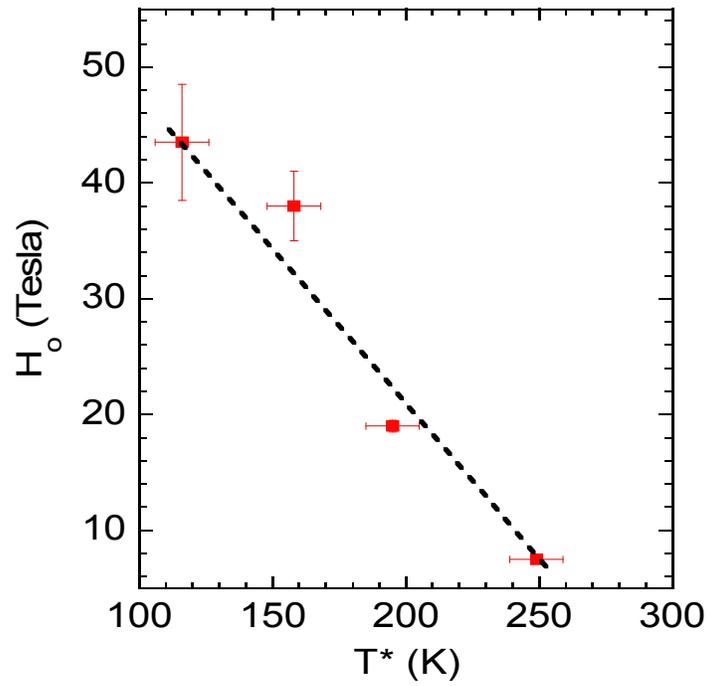